\documentclass{article}
\usepackage{amssymb}

\usepackage{graphicx}
\usepackage{amsmath}
\usepackage{bbm}
\usepackage{epstopdf}
\usepackage[utf8]{inputenc}
\usepackage[T1]{fontenc}
\usepackage{mathrsfs}

\begin{document}

\title{Quantum Trajectories for Squeezed Input Processes: Explicit Solutions}
\author{Anita D\k{a}browska, John Gough}
\maketitle

\begin{abstract}

We consider the quantum (trajectories) filtering equation for the case when the system is driven by Bose field inputs prepared in an arbitrary non-zero mean Gaussian state. The {\it a posteriori} evolution of the system is conditioned by the results of a single or double homodyne measurements. The system interacting with the Bose field is a single cavity mode taken initially in a Gaussian state. We show explicit solutions using the method of characteristic functions to the filtering equations exploiting the linear Gaussian nature of the problem.

\end{abstract}

\section{Introduction}
The aim of this paper is to write explicite solutions to equations for the quantum trajectories for a linear system (cavity mode) driven by inputs in a general Gaussian state. The quantum filtering theory developed by V. P. Belavkin \cite{Bel80}-\cite{Bel89} describes the dynamics of a dissipative Markovian quantum system under an indirect and continuous in time observation. The model is a quantum analog of the classical filtering theory \cite{Strat59}-\cite{DavisMarcus} with the framework of the quantum stochastic Ito calculus (QSC) \cite{HP} being necessary to capture the physical noise. The Bose field interacting with the system here plays the role of measuring apparatus. The evolution of the system conditioned by the results of the measurement is given by the filtering equation, and the conditional density matrix depending on the random results of the measurement can be viewed as a stochastic process. The theory of quantum filtering, or quantum trajectories as it is known in the physics literature, has been developed and is used nowadays in a technique of quantum measurement and control \cite{Bar90}-\cite{hofer15}.

The Belavkin filtering equation, also known as the stochastic master equation, was derived for the Bose field in a pure Gaussian state such as a vacuum \cite{Bel92a}, squeezed vacuum \cite{BouGutMaa04}, and coherent state \cite{GouKos10, DabSta011}. The rigorous derivations of the quantum filter in more general case when the systems is coupled to the multiple fields in arbitrary zero-mean jointly Gaussian state was given in \cite{Nur14}. The filtering equation for the case when the output field mixed with the field in a Gaussian state one can find in \cite{DabGou14}. In the latter we presented the result for arbitrary zero-mean and nonzero-mean Gaussian state of the Bose field including vacuum, squeezed vacuum, coherent, thermal, squeezed thermal and squeezed pure state. Recently the quantum filtering theory has been also expanded to the non-classical states of the Bose field such as a single photon states \cite{GouJamNur12} and a superposition of continuous-mode coherent states \cite{GouJamNur13}. We mention that recent experiments have shown that single photon homodyning may become a realistic procedure
\cite{Furusawa15}.

In this paper we describe the {\it a posteriori} evolution of a harmonic oscillator (an optical cavity mode) driven by the Bose field in a Gaussian state for two measurement schemes: a single homodyne measurement and a double homodyne measurement. Let us remind that in the first of these schemes we measure only one quadrature of the system while the second one allows us inaccurate joint measurement of two quadratures of the system. In the paper we provide the reader with the differential equations for the {\it a posteriori} mean values of the system's operators, show the condition for preservation of the purity of the system's state and assuming that a cavity mode is initially in a Gaussian state we give an analytical solutions to the quantum filter and compare them with the results for {\it a priori} dynamics.

\section{Quantum stochastic processes}

Let us consider two independent annihilation processes $B_{j}\left( \cdot \right) $ $\left( j=1,2\right)$ and related to them two independent creation processes $B_{j}\left( \cdot \right)^{\ast}$ $\left( j=1,2\right)$ satisfying the  
commutation relations \cite{HL85}-\cite{GarZol00}
\begin{equation*}
\left[ B_{j}\left( t\right) ,B_{k}\left( s\right) \right]\;=\;[B_{j}(t)^{\ast},B_{k}
(s)^{\ast}] \;=\;0\,,\;\;\;\;\left[ B_{j}\left( t\right) ,B_{k}\left( s\right)^{\ast } \right]\;=\;\delta
_{jk} \left(t\wedge s\right)\,,
\end{equation*}
where $t \wedge s=\mathrm{min}(t,s)$. The processes  $B_{j}\left( \cdot \right) $ $\left( j=1,2\right)$,  $B_{j}\left( \cdot \right)^{\ast} $ $\left( j=1,2\right)$ are quantum analogues of classical Wiener process. 

We analyze the case when the baths are in Gaussian states 
with the quantum mean values
\begin{equation*}
\mathbb{E}\left[ B_{j}\left( t\right) B_{k}\left( s\right) \right] = \delta_{jk} m_{j}\left(t\wedge s\right)\,,
\end{equation*}
\begin{equation*}
\mathbb{E}\left[ B_{j}\left( t\right)^{\ast} B_{k}\left( s\right) \right] = \delta_{jk} n_{j}\left(t\wedge s\right)\,,
\end{equation*}
\begin{equation*}
\mathbb{E}\left[ B_{j}\left( t\right) B_{k}\left( s\right)^{\ast} \right] = \delta_{jk} (n_{j}+1)\left(t\wedge s\right)\,,
\end{equation*}
\begin{equation*}
\mathbb{E}\left[ B_{j}\left( t\right) \right] =\beta_{j} t\,\;\;\; \left( j=1,2\right)\,.
\end{equation*}
The It\={o} multiplication tables are given then by
\begin{equation}\label{tables}
 \begin{tabular}{l|ll}
$\times $ & $dB_{j}(t)$ & $dB_{j}(t)^{\ast}$ \\ \hline
$dB_{j}(t)$ & $m_{j}dt$ & $\left(n_{j}+1\right) dt$ \\ 
$dB_{j}(t)^{\ast }$ & $n_{j}dt$ & $m^{\ast }_{j}dt$%
\end{tabular},
\end{equation}
where $n_{j}\geq 0$, $\left| m_{j}\right| ^{2}\leq n_{j}\left( n_{j}+1\right)$. All the other products, including $dB_{1}(t)$, $dB_{2}(t)^{\ast}$, etc., vanish. 

When $\beta_{j}=0$, $m_{j}=0$, $n_{j}=\left(e^{\hbar\omega_{j}/k_{B}T}-1\right)^{-1}$, where $\omega_{j}$ is the carrier frequency of the $j$ field, we get a thermal bath. In general, the processes $B_k$ may be represented by an Araki-Woods double-Fock space construction, as done
explicitly in \cite{HL85}. However, in the degenerate case where we have $\left| m_{j}\right| ^{2}= n_{j}\left( n_{j}+1\right)$ 
one can use for each $k$ the representation $B_k \equiv \sqrt{n+1} A_k + \sqrt{n} A^\ast_k$ where $A_k$ is the annihilation process on a single copy of Fock space, with the choice of Fock vacuum. For a finite number of states, the latter case corresponds to Gaussian pure states (vacuum, squeezed vacuum or squeezed coherent state) \cite{BP02, WisMil10}.    

\section{Interaction between observed system and measuring apparatus}

We assume that only the first input, $B_{1}(t)$, interacts with 
the quantum system in interest (system $\mathcal{S}$) and
the evolution of the compound system is described by 
the unitary operator, $U(t)$, which satisfies 
the quantum stochastic differential equation (QSDE) 
of the form \cite{HP, GarZol00}
\begin{equation*}
dU(t)\;=\;\left\{LdB_{1}(t)^{\ast}-L^{\ast}dB_{1}(t)-\left(K+iH\right)dt\right\}
U(t)\,,\;\;\; U(0)\;=\;I\,,
\end{equation*}
where
\begin{equation*}
K\;=\;\frac{1}{2}\left(n_{1}+1\right)L^{\ast}L+\frac{1}{2}n_{1}LL^{\ast}-
\frac{1}{2}m_{1}L^{\ast 2}-\frac{1}{2}m^{\ast}_{1}L^2\,,
\end{equation*}
$H$ is the Hamiltonian of $\mathcal{S}$, and $L$ is a coupling operator. 

The above equation is written in an interaction picture rotating at the frequency $\omega_{1}$ and $H$ is the only part of the Hamiltonian not removed by this procedure.

For any observable $X$ of the system $\mathcal{S}$ in the Heisenberg picture 
\begin{equation*}
j_{t}(X)\;=\;U(t)^{\ast}\left(X\otimes I\right)U(t)\,,
\end{equation*}
we obtain the QSDE \cite{Bar86}
\begin{eqnarray}\label{flow}
dj_{t}(X)&=&\{-ij_{t}\left( \left[ X,H\right] \right) \nonumber
\\&&+\frac{1}{2}\left(n_{1}+1\right)j_{t} 
\left(L^{\ast }\left[ X,L\right]+ \left[L^{\ast },X\right] L \right)
\nonumber\\
&&+\frac{1}{2}n_{1}\, j_{t}\left(L\left[
X,L^{\ast }\right]+ \left[ L,X\right] L^{\ast } \right)\nonumber \\
&&-\frac{1}{2}m^{\ast }_{1}\, j_{t}\left( L\left[ X,L
\right]+\left[ L,X\right] L \right) \nonumber\\
&&-\frac{1}{2}m_{1}\, j_{t}\left( L^{\ast }\left[ X,L^{\ast }\right]+
\left[ L^{\ast },X\right] L^{\ast
} \right)\} \,dt \nonumber\\
&&+j_{t}\left(\left[X,L\right]\right)\,dB_{1}(t)^{\ast}
+j_{t}\left(\left[L^{\ast},X\right]\right)\,dB_{1}(t)\,,
\end{eqnarray}
which one can check it by making use of the rules of QSC and the multiplication table (\ref{tables}).

Let us remind that according to the interpretation of the model given by Gardiner and Collet \cite{GarCol85} the operator $B_{1}(t)$, $B_{1}(t)^{\ast}$ refer to the Bose field before the interaction with the system $\mathcal{S}$, while the field 
after the interaction is described by the output processes 
\begin{equation*}
B_{1}^{\text{out}}(t)\;=\;U(t)^{\ast}\left(I\otimes B_{1}(t)\right)U(t)\,,
\end{equation*}
\begin{equation*}
B_{1}^{\text{out}}(t)^{\ast}\;=\;U(t)^{\ast}\left(I\otimes B_{1}(t)^{\ast}\right)U(t)\,.
\end{equation*}
Thus the output field is the field in the Heisenberg picture and we have \cite{Bar06}
\begin{equation*}
B_{1}^{\text{out}}(t)\;=\;\int_{0}^{t}\left(j_{s}(L)ds+dB_{1}(s)\right)\,,
\end{equation*}
\begin{equation*}
B_{1}^{\text{out}}(t)^{\ast}\;=\;\int_{0}^{t}\left(j_{s}(L^{\ast})ds
+dB_{1}(s)^{\ast}\right)\,.
\end{equation*}
The output field conveys information about the system $\mathcal{S}$. Note that the increments $dB_{1}^{\text{out}}(t)$, $dB_{1}^{\text{out}}(t)^{\ast}$ satisfy the multiplication table of the form (\ref{tables}). Thus the output
field remain Bose free field.

\section{Posterior dynamics for a single homodyne observation}

In this section we shall describe an evolution of a single-mode field 
in a cavity conditioned by the results of a continuous in time measurement of 
an optical quadrature when the cavity is driven by a vacuum, coherent, thermal 
or squeezed input. 

We take the hamiltonian $H=\hbar\delta a^{\ast}a$, where $\delta=\omega_{c}-\omega_{1}$, $\omega_{c}$ is the frequency of the cavity mode, and $a$ stands for annihilation operator, and the coupling operator $L=\sqrt{\mu}a$ with $\mu \in \mathbb{R}_{+}$. 

Let us consider the continuous measurement of the output process
\begin{equation}\label{outprocess}
Y(t)\;=\;e^{i\theta}B^{\text{out}}_{1}(t)+e^{-i\theta}B^{\text{out} }_{1}(t)^{\ast}\,,
\end{equation}
where $\theta\in [0,2\pi)$. Observation of $Y(t)$ can be interpreted as the indirect and imperfect measurement of the cavity mode observable $e^{i\theta}a+e^{-i\theta}a^{\ast}$.  

A measurement of phase-dependent quantities is possible by making use of the homodyne/heterodyne detection scheme where a signal escaping from a cavity is superposed with an auxiliary field (a local oscillator) using a beam splitter \cite{Bar06, GarZol00}. In practise, the measurement of $Y(t)$ is realizable by an implementation of the homodyne scheme. We get the homodyne measurement if the carrier frequency of the input field is equel to the carrier frequency of the local oscillator. When this condition is not satisfied, we have the case of the heterodyne measurement. 

The density matrix of the cavity mode under the continuous observation of the process $Y(t)$ satisfies the stochastic filtering equantion of the form \cite{DabGou14}
\begin{eqnarray}\label{filter1}
d\varrho _{t} &=&\bigg\{
-i\left[\delta a^{\ast}a,\varrho_{t}\right]+\sqrt{\mu}\left[a,\varrho_{t}\right]\,\beta_{1}^{\ast}
+\sqrt{\mu}\left[\varrho_{t},a^{\ast}\right]\,\beta_{1}\,\nonumber
\\&&+\frac{\mu}{2}\left(n_{1}+1\right) 
\left(\left[a\varrho_{t},a^{\ast}\right]+\left[a,
\varrho_{t}a^{\ast}\right]  \right)\nonumber \\
&&+\frac{\mu}{2}n_{1}\left( \left[a^{\ast}\varrho_{t},a\right]+\left[a^{\ast},
\varrho_{t}a\right] \right) \,\nonumber \\
&&-\frac{\mu}{2}m^{\ast }_{1}
\left( \left[ a\varrho_{t},a\right]+\left[a, \varrho_{t}a
\right] \right) \,\nonumber \\
&&-\frac{\mu}{2}m_{1}\left( \left[a^{\ast }\varrho_{t},a^{\ast}\right] +\left[a^{\ast},\varrho_{t}a^{\ast}\right] \right)\bigg\} \,dt\nonumber \\
&&+\frac{\sqrt{\mu}}{\kappa}\bigg\{e^{i\theta}a\varrho_{t}+e^{-i\theta}
\varrho_{t}a^{\ast }-\left(e^{i\theta}\langle a\rangle_{t}+e^{-i\theta}\langle a^{\ast}\rangle_{t}\right)\varrho_{t}\nonumber \\
&& +\left( e^{i\theta}n_{1}+e^{-i\theta}m^{\ast }_{1}\right) 
[a,\varrho_{t}]\nonumber\\
&&+\left( e^{-i\theta}n_{1}+e^{i\theta}m_{1}\right)[\varrho_{t},a^{\ast
}] \bigg\} d\tilde{Y}\left( t\right)\,,
\end{eqnarray}
where
\begin{eqnarray*}
d\tilde{Y}(t)&=&dY(t)-\sqrt{\mu}\left(e^{i\theta}\langle a\rangle_{t}+e^{-i\theta}
\langle a^{\ast}\rangle_{t}\right)dt\\
&&-\left(e^{i\theta}\beta_{1}
+e^{-i\theta}\beta_{1}^{\ast}\right)dt\,,
\end{eqnarray*}
$\kappa=1+2n_{1}+e^{2i\theta}m_{1}+e^{-2i\theta}m^{\ast}_{1}$, and $\langle .\rangle_{t}=\mathrm{Tr}(\varrho_{t}.)$ is the {\it a posteriori} mean value of an operator. The above equation describes the condition evolution of the cavity mode with respect to the commutative von Neumann algebra $Y^t=\{Y\left(s\right)|s\leq t\}$. Note that the mean value of $d\tilde{Y}(t)$ is zero, $(d\tilde{Y}(t))^2=\kappa dt$ and the process  $\tilde{Y}(t)/\sqrt{\kappa}=\int_{0}^{t}d\tilde{Y}(s)/\sqrt{\kappa}$ is izometric to the standard Wiener process. 

One can check that if $|m_{1}|^2=n_{1}(n_{1}+1)$, Eq. (\ref{filter1}) transforms pure states into pure states and it is equivalent to the stochastic filtering equation  
\begin{eqnarray*}
d|\psi_{t}\rangle&=& \big\{-i\delta a^{\ast}a-\frac{\mu}{2}\left(n_{1}+1\right)a^{\ast}a-\frac{\mu}{2}n_{1}aa^{\ast}
\nonumber\\
&&+
\frac{\mu}{2}m_{1}(a^{\ast})^{2}+\frac{\mu}{2}m^{\ast}_{1}a^2+
\sqrt{\mu}\beta_{1}^{\ast}a-\sqrt{\mu}\beta_{1}a^{\ast}\nonumber\\
&&+\frac{\mu e^{-i\theta}}{\kappa}\langle a^{\ast}\rangle_{t}\left[a\left(e^{i\theta}n_{1}
+e^{-i\theta}m_{1}^{\ast}+e^{i\theta}\right)\right.\nonumber\\
&&\left.-a^{\ast}\left(e^{i\theta}m_{1}
+e^{-i\theta}n_{1}\right)\right]-\frac{\mu}{2\kappa}\langle a\rangle_{t}\langle a^{\ast}\rangle_{t}
\big\}|\psi_{t}\rangle\, dt\nonumber\\
&&+\frac{\sqrt{\mu}}{\kappa}\left[a\left(e^{i\theta}n_{1}+e^{-i\theta}m_{1}^{\ast}+
e^{i\theta}\right)
\right.\nonumber \\
&&-\left.a^{\ast}\left(e^{i\theta} m_{1}+e^{-i\theta}n_{1}\right)-e^{i\theta}\langle a\rangle_{t}\right]|\psi_{t}\rangle\, d\tilde{Y}(t)\,,
\end{eqnarray*}
for the {\it a posteriori} wave function $|\psi_{t} \rangle$. Here $\langle .\rangle_{t}=\langle \psi_{t}|(.)\psi_{t}\rangle$ and initial condition states as $|\psi_{t=0}\rangle=|\psi_{0}\rangle$. Thus we obtain preservation of state purity when the field driving the system is pure. The validity of the above equation one can check by differentiating $\varrho_{t}=|\psi_{t}\rangle\langle \psi_{t}|$.

To solve Eq. (\ref{filter1}), we will use the normal ordered characteristic function associated with the state:
\begin{equation}\label{trans}
\tilde{\varrho}(\xi^{\ast},\xi)\; =\;\mathrm{Tr}\left(e^{-i\xi^{\ast}a}\varrho\, e^{-i\xi a^{\ast}}\right)
\end{equation}
with the inverse given by
\begin{equation*}
\varrho\;=\; \frac{1}{\pi}\int d^{2}\xi\, e^{i\xi^{\ast}a}\, e^{i\xi a^{\ast}}\tilde{\varrho}(\xi^{\ast},\xi)\,.
\end{equation*}
Note that $\tilde{\rho}$ is related to the $P$-function in quantum optics \cite{GarZol00}: here the state is represented in terms of coherent states as $\rho \equiv \int |\alpha \rangle \langle \alpha | \, P(\alpha ) d^2 \alpha$ and $P(\alpha ) = \frac{1}{\pi^2} \int e^{i \xi^\ast \alpha +i \xi \alpha^\ast} \tilde{\rho} (\xi^\ast , \xi ) \, d^2 \xi$.

Making use of the transformation (\ref{trans}), one can convert Eq. (\ref{filter1}) into the form
\begin{eqnarray}\label{filter2}
d\tilde{\varrho}(\xi^{\ast},\xi; t)&=&
\left\{-\left(i\delta+\frac{\mu}{2}\right)\xi^{\ast}\partial_{\xi^{\ast}}
-\left(-i\delta+\frac{\mu}{2}\right)\xi\partial_{\xi}\right.\nonumber\\
&&+\sqrt{\mu}\beta_{1}i\xi^{\ast}+\sqrt{\mu}\beta_{1}^{\ast}i\xi
-\mu n_{1}\left|\xi\right|^2\nonumber\\
&&\left.-\frac{\mu m_{1}}{2}\xi^{\ast 2} -\frac{\mu m^{\ast}_{1}}{2}\xi^{2}\right\}\tilde{\varrho}(\xi^{\ast},\xi;t)dt\nonumber\\
&&+\frac{\sqrt{\mu}}{\kappa}
\left[e^{i\theta}i\partial_{\xi^{\ast}} +e^{-i\theta}i\partial_{\xi}+i\xi^{\ast}\left(e^{i\theta}m_{1}+e^{-i\theta }n_{1}\right)\right.\nonumber\\
&&\left.\!+i\xi\left(e^{-i\theta}m_{1}^{\ast}\!+\!e^{i\theta}n_{1}\right)
\!-\!e^{i\theta}\langle a\rangle_{t}\!-\!e^{-i\theta}\langle a^{\ast}\rangle_{t}\right]\tilde{\varrho}(\xi^{\ast},\xi;t)d\tilde{Y}(t),
\end{eqnarray}
where $\partial_{\xi}=\partial/\partial_{\xi}$, $\partial_{\xi^{*}}=\partial/\partial_{\xi^{*}}$.

We assume that the cavity mode is initially in the Gaussian state
\begin{equation}\label{gauss}
\tilde{\rho}\left(\xi^{*},\xi,t=0\right)\;=\;\exp\left[
-i\left(\xi^{\ast}\alpha_{0}+\xi \alpha_{0}^{\ast}\right)-
\frac{1}{2}\left(\xi^{\ast 2}\zeta_{0}+\xi^{2}\zeta_{0}^{\ast}\right)
-|\xi|^{2}\nu_{0}\right]\,,
\end{equation}
where $\alpha_{0}, \zeta_{0} \in \mathbb{C}$, $\nu_{0}\geq 0$. One can easily check that   
\begin{equation*}
\mathrm{Tr}[a\rho_{t=0}]=\alpha_{0}\,,\;\;\;\mathrm{Tr}[a^2\rho_{t=0}]=\zeta_{0}+\alpha^2_{0}\,,\;\;\;
\mathrm{Tr}[a^{\ast}a\rho_{t=0}]=\nu_{0}+|\alpha_{0}|^2\,.
\end{equation*}

We shall prove that the solution to Eq. (\ref{filter2}) corresponding to the initial state (\ref{gauss}) has the Gaussian form
\begin{equation}
\tilde{\rho}(\xi^{\ast},\xi;t )\!=\!\exp\left[
-i\left(\xi^{\ast}\langle a\rangle_{t}\!+\!\xi \langle a^{\ast}\rangle_{t}\right)-
\frac{1}{2}\left(\xi^{\ast 2}\zeta(t)\!+\!\xi^{2}\zeta(t)^{\ast}\right)
\!-\!|\xi|^{2}\nu(t)\right]
\end{equation}
where $\langle a\rangle_{t}=\mathrm{Tr}[a \rho_{t}]$. 

To this end, it is convenient to rewrite Eq. (\ref{filter2}) in terms of the stochastic function 
\begin{equation}\label{sol}
l(\xi^{\ast},\xi;t)\;=\;-\ln\tilde{\rho}(\xi^{\ast},\xi;t)\,.
\end{equation}
Using the It\={o} formula
\begin{equation*}
dl\;=\;-\frac{1}{\tilde{\rho}}d \tilde{\rho}+\frac{1}{2 (\tilde{\rho})^2} \left(d\tilde{\rho}\right)^2\,,
\end{equation*}
and the tables (\ref{tables}), we obtain
\begin{eqnarray}\label{filter12}
dl&=&
\left\{-\left(i\delta+\frac{\mu}{2}\right)\xi^{\ast}\partial_{\xi^{\ast}}\,l
-\left(-i\delta+\frac{\mu}{2}\right)\xi\partial_{\xi}\,l\right.\nonumber\\
&&-\sqrt{\mu}\beta_{1}i\xi^{\ast}-\sqrt{\mu}\beta_{1}^{\ast}i\xi
+\mu n_{1}\left|\xi\right|^2\nonumber\\
&&\left.+\frac{\mu m_{1}}{2}\xi^{\ast 2}+\frac{\mu m^{\ast}_{1}}{2}\xi^{2}\right\}dt\nonumber\\
&&+\frac{\mu}{2\kappa}
\left[e^{i\theta}i\partial_{\xi^{\ast}}\,l +e^{-i\theta}i\partial_{\xi}\,l-i\xi^{\ast}\left(e^{i\theta}m_{1}+e^{-i\theta }n_{1}\right)\right.\nonumber\\
&&\left.-i\xi\left(e^{-i\theta}m_{1}^{\ast}+e^{i\theta}n_{1}\right)
+e^{i\theta}\langle a\rangle_{t}+e^{-i\theta}\langle a^{\ast}\rangle_{t}\right]^2dt\nonumber\\
&&+\frac{\sqrt{\mu}}{\kappa}
\left[e^{i\theta}i\partial_{\xi^{\ast}}\,l +e^{-i\theta}i\partial_{\xi}\,l-i\xi^{\ast}\left(e^{i\theta}m_{1}+e^{-i\theta }n_{1}\right)\right.\nonumber\\
&&\left.-i\xi\left(e^{-i\theta}m_{1}^{\ast}+e^{i\theta}n_{1}\right)
+e^{i\theta}\langle a\rangle_{t}+e^{-i\theta}\langle a^{\ast}\rangle_{t}\right]d\tilde{Y}(t)\,.
\end{eqnarray}
And finally inserting (\ref{sol}) into Eq. (\ref{filter12}) and equating expressions of the same power of parameters $\xi$ and $\xi^{*}$ provides us with the consistent set of the differential equations: 
\begin{eqnarray}\label{apost}
d\langle a\rangle_{t}&=&-\left(i\delta+\frac{\mu}{2}\right)\langle a\rangle_{t}dt-\sqrt{\mu}\beta_{1}dt\nonumber\\
&&+\frac{\sqrt{\mu}}{\kappa}\left[e^{i\theta}\left(\zeta(t)-m_{1}\right)
+e^{-i\theta}\left(\nu(t)-n_{1}\right)\right]d\tilde{Y}(t)\,,
\end{eqnarray}
\begin{eqnarray}\label{dzeta}
\dot{\zeta}(t)&=&-2i\delta\zeta(t)-\mu\left(\zeta(t)-m_{1}\right)\nonumber\\
&&-\frac{\mu}{\kappa}
\left[e^{i\theta}\left(\zeta(t)-m_{1}\right)+e^{-i\theta}\left(\nu(t)-n_{1}\right)\right]^2\,,
\end{eqnarray}
and
\begin{equation}\label{dnu}
\dot{\nu}(t)\;=\;-\mu\left(\nu(t)-n_{1}\right)-\frac{\mu}{\kappa}
\left|e^{i\theta}\left(\zeta(t)-m_{1}\right)+e^{-i\theta}\left(\nu(t)-n_{1}\right)\right|^2
\end{equation}
with the initial conditions: $\langle a\rangle_{t=0}=\alpha_{0}$, $\zeta(t=0)=\zeta_{0}$, and $\nu(t=0)=\nu_{0}$; which completes the proof. 

One can see from Eqs. (\ref{apost}--\ref{dnu}) that the coherent state of the cavity mode is preserved under the observation of $Y(t)$ only when the input field is taken in a vacuum or coherent state. If $m_{1}\neq 0$, then the {\it a posteriori} state becomes squeezed. In the resonance case when $\delta=0$, we obtain the stationary asymptotic solutions of the form: $\langle a\rangle_{\infty}=\frac{-2\beta_{1}}{\sqrt{\mu}}$, ${\zeta}(\infty)=m_{1}$, and
${\nu}(\infty)=n_{1}$. So the system approaches the Gaussian state with the parameter of squeezing defined by the state of the input field. When $m_{1}=0$ then for any value of $\delta$ we have ${\zeta}(\infty)=0$ and
${\nu}(\infty)=n_{1}$.

We may draw from Eqs. (\ref{apost}--\ref{dnu}) also the conclusion that the {\it a posteriori} mean values of optical quadratures $X=a+a^{\ast}$ and $P=\left(a-a^{\ast}\right)/\mathrm{i}$ of the cavity, in general, depend on the measured noise, whereas their dispersions, given by 
\begin{equation}\label{dispx}
(\Delta X(t))^2\;=\;1+2\nu(t)+2\mathrm{Re}\zeta(t)\,,
\end{equation}
\begin{equation}\label{dispy}
(\Delta P(t))^2\;=\;1+2\nu(t)-2\mathrm{Re}\zeta(t)\,,
\end{equation}
remain non-random. 
 
Let us note that Eqs. (\ref{dzeta}) and (\ref{dnu}) can be written as matrix Riccati differential equation of the form
\begin{equation}\label{Riccati}
\dot{Z}(t)\;=\;Z(t)TZ(t)+RZ(t)+Z(t)R^{\ast}+W\,,
\end{equation}
where 
\begin{equation*}
Z(t)\;=\;\left[\begin{array}{cc} \nu(t) & \zeta(t) \\
\zeta(t)^{\ast} & \nu(t)\end{array}\right]\,, 
\end{equation*}
\begin{equation*}
T\;=\;-\frac{\mu}{\kappa}\left[\begin{array}{cc} 1& e^{-2i\theta} \\
e^{2i\theta} & 1\end{array}\right]\,,
\end{equation*}
\begin{equation*}
R=
\left[\begin{array}{cc} -i\delta  & 0 \\
0 & i\delta\end{array}\right]
+
\frac{\mu}{\kappa} \left[\begin{array}{cc} -  \frac{\kappa}{2}+n_{1}+e^{2i\theta}m_{1}  &   m_{1}+e^{-2i\theta}n_{1}  \\
 m_{1}^{\ast}+e^{2i\theta}n_{1}  &  -\frac{\kappa}{2}+n_{1}+e^{-2i\theta}m_{1}^{\ast}
\end{array}\right],
\end{equation*} 
\begin{equation*}
W\;=\;\frac{\mu}{\kappa} \left[\begin{array}{cc}
 n_{1}+n^{2}_{1}-\left|m_{1}\right|^2  &   m_{1}+e^{-2i\theta}(\left|m_{1}\right|^2-n^{2}_{1})  \\
  m_{1}^{\ast}+
e^{2i\theta}(\left|m_{1}\right|^2-n^{2}_{1})  &   n_{1}+n^{2}_{1}-\left|m_{1}\right|^2 \end{array} \right] ,
\end{equation*} 
with the initial condition $Z(t=0)=\left[\begin{array}{cc} \nu_{0} & \zeta_{0} \\
\zeta^{*}_{0} & \nu_{0}\end{array}\right]$. 

Eq. (\ref{Riccati}) can be solved by using the matrix fraction decomposition technique \cite{AndMoo71} which consists in representing matrix $Z(t)$ in the form
\begin{equation*}
Z(t)\;=\;Z_{1}(t)Z^{-1}_{2}(t)\,.
\end{equation*}
The initial values of $Z_{1}(t)$ and $Z_{2}(t)$ are constrained by the initial value of $Z(t)$. One can check that the differential equations for the numerator and denominator matrices are linear and they can be read as follows
\begin{equation*}
\left[\begin{array}{c}
\dot{Z}_{1}(t) \\
\dot{Z}_{2}(t)\end{array}\right]\;=\;
\left[\begin{array}{cc}
R & W \\
-T & -R^{\ast}\end{array}\right]
\left[\begin{array}{c}
{Z}_{1}(t) \\
{Z}_{2}(t)\end{array}\right]\,.
\end{equation*} 
The solution to the foregoing can be written as the product
\begin{equation}\label{solriccati}
\left[\begin{array}{c}
Z_{1}(t) \\
Z_{2}(t)\end{array}\right]\;=\;
e^{\Upsilon t}
\left[\begin{array}{c}
Z(0) \\
\mathbbm{1}\end{array}\right]\,,
\end{equation} 
where
$\Upsilon\;=\;\left[\begin{array}{cc}
R & W \\
-T & -R^{\ast}\end{array}\right]$ 
is the time-invariant 4-by-4 matrix and $\mathbbm{1}$ is identity matrix of size 2.

\section{Posterior dynamics for a double homodyne observation}

Joint perfect measurement of two optical quadratures of a quantum system is not possible, but one can analyze their simultaneous imperfect measurement. Using a double homodyne scheme one can mix the field escaping from the cavity (the output field) with an auxiliary field and measure simultaneously the two signals:  
 \begin{equation*}
Y_{1}(t)\;=\;\frac{1}{\sqrt{2}}\left(Q_{1}(t)+Q_{2}(t)\right)\,,
\end{equation*}
\begin{equation*}
Y_{2}(t)\;=\;\frac{1}{\sqrt{2}}\left(P_{1}(t)-P_{2}(t)\right)\,,
\end{equation*}
where $Q_{1}(t)=B^{\text{out}}_{1}(t)+B^{\text{out}}_{1}(t)^{\ast}$, $P_{1}(t)=\left(B^{\text{out}}_{1}(t)-B^{\text{out}}_{1}(t)^{\ast}\right)/i$ are quadratures of the field escaping from the cavity and $Q_{2}(t)=B_{2}(t)+B_{2}(t)^{\ast}$, $P_{2}(t)=\left(B_{2}(t)-B_{2}(t)^{\ast}\right)/i$ describe the squeezed noise field do not interacting with $\mathcal{S}$. The observed processes $Y_{1}(t)$ and $Y_{2}(t)$ satisfy the condition $[Y_{1}(t),Y_{2}(s)]=0$ for all $ t,s \geq 0$ and their increments 
\begin{eqnarray*}
dY_{1}\left( t\right)  &=&\frac{1}{\sqrt{2}}\left\{
dB_{1}\left( t\right) +\sqrt{\mu}\,j_{t}\left( a\right) dt+dB_{2}\left( t\right) +\text{%
H.c.}\right\} , \\
dY_{2}\left( t\right)  &=&\frac{1}{\sqrt{2}i}\left\{dB_{1}\left( t\right) +\sqrt{\mu}\,j_{t}\left( a\right) dt-dB_{2}\left( t\right) -\text{%
H.c.}\right\}\,,
\end{eqnarray*}
have non-trivial correlation expressed through the table
\begin{equation*}
\begin{tabular}{l|ll}
$\times $ & $dY_{1}$ & $dY_{2}$ \\ \hline
$dY_{1}$ & $\left( 1+n_{1}+n_{2}+m^{\prime }_{1}+m^{\prime}_{2}\right) dt$ & $\left(m^{\prime\prime}_{1}-m^{\prime \prime }_{2}\right)dt$ \\ 
$dY_{2}$ & $\left(m^{\prime\prime}_{1}-m^{\prime \prime }_{2}\right)dt$ & $\left( 1+n_{1}+n_{2}-m^{\prime}_{1}-m_{2}^{\prime}\right) dt
$%
\end{tabular}
\end{equation*}
where $m_{1}=m^{\prime }_{1}+im^{\prime \prime }_{1}$,  $m_{2}=m^{\prime }_{2}+im^{\prime \prime }_{2}$ are the decompositions of the
squeezing parameters into their real and imaginary parts.

The filtering equation for the {\it a posteriori} state, $\varrho_{t}$, takes in this case the form \cite{DabGou14}
\begin{eqnarray}\label{filter3}
d\varrho _{t} &=&
\bigg\{-i\left[\delta a^{\ast}a,\varrho_{t}\right]+\sqrt{\mu}\left[a,\varrho_{t}\right]\,\beta_{1}^{\ast}
+\sqrt{\mu}\left[\varrho_{t},a^{\ast}\right]\,\beta_{1}\nonumber
\\&&+\frac{\mu}{2}\left(n_{1}+1\right) 
\left(\left[a\varrho_{t},a^{\ast}\right]+\left[a,
\varrho_{t}a^{\ast}\right]  \right)\nonumber \\
&&\quad +\frac{\mu}{2}n_{1}\left( \left[a^{\ast}\varrho_{t},a\right]+\left[a^{\ast},
\varrho_{t}a\right] \right)\nonumber \\
&&\quad -\frac{\mu}{2}m^{\ast }_{1}\left( \left[ a\varrho_{t},a\right]+\left[a, \varrho_{t}a
\right] \right)\nonumber \\
&&\quad -\frac{\mu}{2}m_{1}\left( \left[a^{\ast }\varrho_{t},a^{\ast}\right] +\left[a^{\ast},\varrho_{t}a^{\ast}\right] \right)\bigg\}\,dt\nonumber \\
&&  +\frac{\sqrt{2\mu}}{2\Delta }
\bigg\{ \left(1+n_{1}+n_{2}-m_{1}^{\prime}-m_{2}^{\prime}\right)
\mathcal{J}_{1}(\varrho_{t}) \nonumber\\
&& \quad +i\left(m_{1}^{\prime\prime}-m_{2}^{\prime\prime}\right)
\mathcal{J}_{2}(\varrho_{t})
\bigg\} d\tilde{Y}_{1}\left( t\right)\nonumber\\
&&+\frac{\sqrt{2 \mu}}{2\Delta i }
\bigg\{\left(1+n_{1}+n_{2}+m_{1}^{\prime}+m_{2}^{\prime}\right)
\mathcal{J}_{2}(\varrho_{t}) \nonumber\\
&&\quad -i\left(m_{1}^{\prime\prime}-m_{2}^{\prime\prime}\right)
\mathcal{J}_{1}(\varrho_{t})
\bigg\} d\tilde{Y}_{2}\left( t\right)\,,
\end{eqnarray}
where
\begin{eqnarray*}
\mathcal{J}_{1}(\varrho_{t})&=&a\varrho_{t}+
a^{\ast }\varrho_{t}-\langle a\rangle_{t}\varrho_{t}-
\langle a^{\ast}\rangle_{t}\varrho_{t} \\
&&+\left(n_{1}+m^{\ast}_{1}\right) 
[a,\varrho_{t}]+\left(n_{1}+m_{1}+1\right)
[\varrho_{t},a^{\ast
}]\,,
\end{eqnarray*}
\begin{eqnarray*}
\mathcal{J}_{2}(\varrho_{t})&=&a\varrho_{t}-
a^{\ast }\varrho_{t}-\langle a\rangle_{t}\varrho_{t}+
\langle a^{\ast}\rangle_{t}\varrho_{t} \\
&&+\left(n_{1}-m^{\ast}_{1}\right) 
[a,\varrho_{t}]-\left(n_{1}-m_{1}+1\right)
[\varrho_{t},a^{\ast
}]\,,
\end{eqnarray*}
\begin{equation*}
d\tilde{Y}_{1}(t)\;=\;dY_{1}(t)-\sqrt{\frac{\mu}{2}}\left(\langle a\rangle_{t}+\langle a^{\ast}\rangle_{t}\right)dt-\sqrt{2}\left(\mathrm{Re}\,\beta_{1}
+\mathrm{Re}\,\beta_{2}\right)dt\,,
\end{equation*}
\begin{equation*}
d\tilde{Y}_{2}(t)\;=\;dY_{2}(t)+\sqrt{\frac{\mu}{2}}i\left(\langle a\rangle_{t}-\langle a^{\ast}\rangle_{t}\right)dt-\sqrt{2}\left(\mathrm{Im}\,\beta_{1}
-\mathrm{Im}\,\beta_{2}\right)dt\,,
\end{equation*}
and
\begin{equation}
\Delta=\left(1+n_{1}+n_{2}\right)^2-\left(m_{1}^{\prime }+m_{2}^{\prime}\right)^2-\left(m_{1}^{\prime\prime}-m_{2}^{\prime \prime}\right)^2\,.
\end{equation}

Here the condition $\left(\rho_{t+dt}\right)^2=\rho_{t+dt}$ if $\left(\rho_{t}\right)^2=\rho_{t}$ is satisfied if and only if $|m_{1}|^2=n_{1}(n_{1}+1)$ and  $|m_{2}|^2=n_{2}(n_{2}+1)$. It means that the conditional state of the cavity mode preserves its purity if and only if both electromagnetic fields used by us for observing the system are in pure states. For instance, if the first field is in a coherent state ($\beta_{1}\neq 0$, $m_{1}=0$, $n_{1}=0$) and the cavity is initially in a pure state, then Eq. (\ref{filter3}) is equivalent to the stochastic filtering equation 
\begin{eqnarray*}
d|\psi_{t}\rangle&=& \bigg\{-\left(i\delta +\frac{\mu}{2}\right)a^{\ast}a+\mu a\langle a^{\ast}\rangle_{t}-\frac{\mu}{2}\langle a\rangle_{t}\langle a^{\ast}\rangle_{t}\nonumber\\
&& \qquad +\sqrt{\mu}\left(\beta_{1}^{\ast}a-
\beta_{1}a^{\ast}\right)\bigg\}|\psi_{t}\rangle\,dt\nonumber\\
&&+\frac{\sqrt{\mu}}{\sqrt{2}\Delta}\bigg\{ \left(1+n_{2}-m_{2}\right) d\tilde{Y}_{1}(t) \nonumber\\
&& \qquad -i\left(1+n_{2}+m_{2}\right)d\tilde{Y}_{2}(t)\bigg\} \left(a-\langle a\rangle_{t}\right)|\psi_{t}\rangle\,
\end{eqnarray*}
for the {\it a posteriori} wavefunction $|\psi_{t}\rangle$.

We shall prove that Eq. (\ref{filter3}) transforms Gaussian states into Gaussian states. Using (\ref{trans}), we obtain from (\ref{filter3}) the stochastic equation 
\begin{eqnarray}\label{filter4}
d\tilde{\varrho}(\xi^{\ast},\xi; t)&=&
\left\{-\left(i\delta+\frac{\mu}{2}\right)\xi^{\ast}\partial_{\xi^{\ast}}
-\left(-i\delta+\frac{\mu}{2}\right)\xi\partial_{\xi}\right.\nonumber\\
&&+\sqrt{\mu}\beta_{1}i\xi^{\ast}+\sqrt{\mu}\beta_{1}^{\ast}i\xi
-\mu n_{1}\left|\xi\right|^2\nonumber\\
&&\left.-\frac{\mu m_{1}}{2}\xi^{\ast 2} -\frac{\mu m^{\ast}_{1}}{2}\xi^{2}\right\}\tilde{\varrho}(\xi^{\ast},\xi;t)dt\nonumber\\
&&+\frac{\sqrt{\mu}}{\sqrt{2}\Delta } \left[i\partial_{\xi^{\ast}}+i\partial_{\xi}+i\xi^{\ast}(m_1+n_1)
\right.\nonumber\\
&&\left.+i\xi(m_{1}^{\ast}+n_{1})-\langle a\rangle_{t}-\langle a^{\ast}\rangle_{t}\right]\tilde{\varrho}(\xi^{\ast},\xi;t)\nonumber\\
&&\times[(1+n_{1}+n_{2}
-m_{1}^{\prime}-m_{2}^{\prime})d\tilde{Y}_{1}(t)\nonumber\\
&&-(m_{1}^{\prime\prime}-m_{2}^{\prime\prime})d\tilde{Y}_{2}(t)]\nonumber\\
&&+\frac{\sqrt{\mu}}{\sqrt{2}\Delta i} \left[i\partial_{\xi^{\ast}}-i\partial_{\xi}+i\xi^{\ast}(m_1-n_1)
\right.\nonumber\\
&&\left.-i\xi(m_{1}^{\ast}-n_{1})-\langle a\rangle_{t}+\langle a^{\ast}\rangle_{t}\right]\tilde{\varrho}(\xi^{\ast},\xi;t)\nonumber\\
&&\times[(1+n_{1}+n_{2}
+m_{1}^{\prime}+m_{2}^{\prime})d\tilde{Y}_{2}(t)\nonumber\\
&&-(m_{1}^{\prime\prime}-m_{2}^{\prime\prime})d\tilde{Y}_{1}(t)]\,.
\end{eqnarray}

In this case the equation for the function $l(t)$ has the form
\begin{eqnarray}\label{filter5}
dl&=&
\left\{-\left(i\delta+\frac{\mu}{2}\right)\xi^{\ast}\partial_{\xi^{\ast}}\,l
-\left(-i\delta+\frac{\mu}{2}\right)\xi\partial_{\xi}\,l\right.\nonumber\\
&&-\sqrt{\mu}\beta_{1}i\xi^{\ast}-\sqrt{\mu}\beta_{1}^{\ast}i\xi
+\mu n_{1}\left|\xi\right|^2\nonumber\\
&&\left.+\frac{\mu m_{1}}{2}\xi^{\ast 2}+\frac{\mu m^{\ast}_{1}}{2}\xi^{2}\right\}dt+\frac{(dl)^2}{2}\nonumber\\
&&+\frac{\sqrt{\mu}}{\sqrt{2}\Delta } \left[i\partial_{\xi^{\ast}}l+i\partial_{\xi}l-i\xi^{\ast}(m_1+n_1)
\right.\nonumber\\
&&\left.-i\xi(m_{1}^{\ast}+n_{1})+\langle a\rangle_{t}+\langle a^{\ast}\rangle_{t}\right]\nonumber\\
&&\times[(1+n_{1}+n_{2}
-m_{1}^{\prime}-m_{2}^{\prime})d\tilde{Y}_{1}(t)\nonumber\\
&&-(m_{1}^{\prime\prime}-m_{2}^{\prime\prime})d\tilde{Y}_{2}(t)]\nonumber\\
&&+\frac{\sqrt{\mu}}{\sqrt{2}\Delta i} \left[i\partial_{\xi^{\ast}}l-i\partial_{\xi}l-i\xi^{\ast}(m_1-n_1)
\right.\nonumber\\
&&\left.+i\xi(m_{1}^{\ast}-n_{1})+\langle a\rangle_{t}-\langle a^{\ast}\rangle_{t}\right]\tilde{\varrho}(\xi^{\ast},\xi;t)\nonumber\\
&&\times[(1+n_{1}+n_{2}
+m_{1}^{\prime}+m_{2}^{\prime})d\tilde{Y}_{2}(t)\nonumber\\
&&-(m_{1}^{\prime\prime}-m_{2}^{\prime\prime})d\tilde{Y}_{1}(t)]\,.
\end{eqnarray}
Now inserting 
\begin{equation*}
l(\xi^{\ast},\xi;t)\;=\;-i\left(\xi^{\ast}\langle a\rangle_{t}\!+\!\xi \langle a^{\ast}\rangle_{t}\right)-
\frac{1}{2}\left(\xi^{\ast 2}\zeta(t)\!+\!\xi^{2}\zeta(t)^{\ast}\right)
\!-\!|\xi|^{2}\nu(t)
\end{equation*}
into Eq. (\ref{filter5}) and evaluating $(dl)^2$ we end up with the consistent set of differential equations:
\begin{eqnarray}\label{apost1}
d\langle a\rangle_{t}&=&-\left(i\delta+\frac{\mu}{2}\right)\langle a\rangle_{t}dt-\sqrt{\mu}\beta_{1}dt\nonumber\\
&&+\frac{\sqrt{\mu}}{\sqrt{2}\Delta}\bigg\{
[(1+n_{1}+n_{2}
-m_{1}^{\prime}-m_{2}^{\prime})d\tilde{Y}_{1}(t)
-(m_{1}^{\prime\prime}-m_{2}^{\prime\prime})d\tilde{Y}_{2}(t)]\nonumber\\
&&\qquad \qquad\times(\zeta-m_{1}+\nu-n_{1})\nonumber\\
&&\qquad \qquad-i[(1+n_{1}+n_{2}
+m_{1}^{\prime}+m_{2}^{\prime})d\tilde{Y}_{2}(t)
-(m_{1}^{\prime\prime}-m_{2}^{\prime\prime})d\tilde{Y}_{1}(t)]\nonumber\\
&&\qquad \qquad\times((\zeta-m_{1})-(\nu-n_{1}))\bigg\}\,,
\end{eqnarray}
\begin{eqnarray}\label{zeta}
\dot{\zeta}(t)&=&-2i\delta\zeta(t)+\mu\left(\zeta(t)-m_{1}\right)\nonumber\\
&&+\frac{\mu}{\Delta}\bigg\{
-2(1+n_{1}+n_{2})(\zeta-m_{1})(\nu-n_{1})\nonumber\\
&&+(m_{1}^{\prime}+m_{2}^{\prime})[(\zeta-m_{1})^2+(\nu-n_{1})^2]\nonumber\\
&&-i(m_{1}^{\prime\prime}-m_{2}^{\prime\prime})[(\zeta-m_{1})^2-(\nu-n_{1})^2]\bigg\}\,,
\end{eqnarray}
\begin{eqnarray}\label{nu}
\dot{\nu}(t)&=&-\mu\left(\nu(t)-n_{1}\right)\nonumber\\
&&+\frac{\mu}{\Delta}\bigg\{
-(1+n_{1}+n_{2})[(\zeta-m_{1})(\zeta^{\ast}-m_{1}^{\ast})+(\nu-n_{1})^2]\nonumber\\
&&+(m_{1}^{\prime}+m_{2}^{\prime})(\nu-n_{1})(\zeta+\zeta^{\ast}-2m_{1}^{\prime})\nonumber\\
&&-i(m_{1}^{\prime\prime}-m_{2}^{\prime\prime})(\nu-n_{1})(\zeta-\zeta^{\ast}-2i m_{1}^{\prime\prime})\bigg\}\,,
\end{eqnarray}
with the initial conditions $\langle a\rangle_{t=0}=\alpha_{0}$, $\zeta(t=0)=\zeta_{0}$, and $\nu(t=0)=\nu_{0}$. 

It is easy to see that if $\delta=0$ we have the asymptotic solutions: $\langle a\rangle_{\infty}=\frac{-2\beta_{1}}{\sqrt{\mu}}$, $\zeta(\infty)=m_{1}$, and $\nu(\infty)=n_{1}$.

For the coherent input Eqs. (\ref{zeta}) and (\ref{nu}) form the homogeneous matrix Riccati differential equation 
\begin{equation}\label{Riccati2}
\dot{Z}(t)\;=\;Z(t)TZ(t)+RZ(t)+Z(t)R^{\ast}\,,
\end{equation}
where 
\begin{equation*}
Z(t)=\left[\begin{array}{cc} \nu(t) & \zeta(t) \\
\zeta(t)^{\ast} & \nu(t)\end{array}\right]\,,
\end{equation*}
\begin{equation*}
T=\frac{\mu}{\Delta}\left[\begin{array}{cc} -\left(1+n_{2}\right) & m^{\ast}_{2} \\
m_{2} & -\left(1+n_{2}\right)\end{array}\right]\,,
\end{equation*}
\begin{equation*}
R=\left[\begin{array}{cc} -\frac{\mu}{2}-i\delta & 0 \\
0 & -\frac{\mu}{2}+i\delta\end{array}\right]\,,
\end{equation*}
and initially $Z(t=0)=\left[\begin{array}{cc} \nu_{0} & \zeta_{0} \\
\zeta^{*}_{0} & \nu_{0}\end{array}\right]$. 

Using (\ref{solriccati}) we obtain the exact expressions of the parameters: 
\begin{equation}\label{zeta2}
\zeta(t)\;=\;\frac{e^{-\left(\mu+2i\delta\right)t}}{D(t)}\left[\zeta_{0}+\frac{\mu m^{\ast}_{2}\left(\nu^{2}_{0}-\left|\zeta_{0}\right|^2\right)}{\Delta\left(\mu -2i\delta\right)}\left(1-e^{-(\mu-2i\delta)t}\right)\right],
\end{equation}
\begin{equation}\label{nu2}
\nu(t)\;=\;\frac{e^{-\mu t}}{D(t)}\left[\nu_{0}+\frac{\left(1+n_{2}\right)\left(\nu^{2}_{0}-\left|\zeta_{0}\right|^2\right)}{\Delta}
\left(1-e^{-\mu t}\right)\right],
\end{equation} 
where
\begin{eqnarray}
D(t)&=&1+\frac{\left(1+n_{2}\right)^2\left(\nu^{2}_{0}-\left|\zeta_{0}\right|^2\right)}{\Delta^2}\left(1-e^{-\mu t}\right)^2+\frac{2\nu_{0}\left(1+n_{2}\right)}{\Delta}\left(1-e^{-\mu t}\right)\nonumber\\
&&-\frac{\mu m^{\ast}_{2}\zeta_{0}^{\ast} }{\Delta\left(\mu-2i\delta\right)}\left(1-e^{-(\mu-2i\delta)t}\right)
-\frac{\mu m_{2} \zeta_{0}}{\Delta\left(\mu+2i\delta\right)}\left(1-e^{-(\mu+2i\delta)t}\right)\nonumber\\
&&-\frac{\mu^2\left|m_{2}\right|^2\left(\nu^{2}_{0}-|\zeta_{0}|^2\right)}
{\Delta^2\left(\mu^2+4\delta^2\right)}
\left(1-e^{-(\mu-2i\delta)t}\right)
\left(1-e^{-(\mu+2i\delta)t}\right).
\end{eqnarray}

Let us notice that for the coherent and vacuum inputs ($m_{1}=m_{2}=0$ and $n_{1}=n_{2}=0$) and the cavity mode being initially in a coherent state, we get $\zeta(t)=0$ and $\nu(t)=0$ for any $t$. Thus, in that case the amplitude $\langle a\rangle_{t}$ is independent of the measured noise. The same conclusion we obtain 
for the conditional evolution under a single homodyne observation.
 
For the large $t$ the parameters $\zeta(t)$ and $\nu(t)$ go to zero and the systems approaches asymptotically coherent state. Therefore after some transient time any information about the initial conditions is lost.

\section{A priori evolution}

In this paragraph we would like to compare the {\it a posteriori} dynamics of the cavity mode with the corresponding {\it a priori} dynamics. A non-selective dynamics of the system we obtain by averaging both sides of Eq. (\ref{filter1}) or respectively (\ref{filter3}) over the outputs. It gives us the equation 
\begin{eqnarray}\label{master1}
\dot{\sigma} _{t} &=&
-i\left[\delta a^{\ast}a,\sigma_{t}\right]+\sqrt{\mu}\left[a,\sigma_{t}\right]\,\beta_{1}^{\ast}
+\sqrt{\mu}\left[\sigma_{t},a^{\ast}\right]\,\beta_{1}\nonumber
\\&&+\frac{\mu}{2}\left(n_{1}+1\right) 
\left(\left[a\sigma_{t},a^{\ast}\right]+\left[a,
\sigma_{t}a^{\ast}\right]  \right)\nonumber \\
&&+\frac{\mu}{2}n_{1}\left( \left[a^{\ast}\sigma_{t},a\right]+\left[a^{\ast},
\sigma_{t}a\right] \right)\nonumber \\
&&-\frac{\mu}{2}m^{\ast }_{1}\left( \left[ a\sigma_{t},a\right]+\left[a, \sigma_{t}a
\right] \right)\nonumber\\
&&-\frac{\mu}{2}m_{1}\left( \left[a^{\ast }\sigma_{t},a^{\ast}\right] +\left[a^{\ast},\sigma_{t}a^{\ast}\right] \right)\,,
\end{eqnarray}
for the {\it a priori} state $\sigma_{t}$. 

Let us remind that this equation does not preserve in general the purity of the state.    

By transformation (\ref{trans}), we get from (\ref{master1}) the following equation
\begin{eqnarray}\label{master2}
\dot{\tilde{\sigma}}(\xi^{\ast},\xi; t)&=&
\left\{-\left(i\delta+\frac{\mu}{2}\right)\xi^{\ast}\partial_{\xi^{\ast}}
-\left(-i\delta+\frac{\mu}{2}\right)\xi\partial_{\xi}\right.\nonumber\\
&&+\sqrt{\mu}\beta_{1}i\xi^{\ast}+\sqrt{\mu}\beta_{1}^{\ast}i\xi
-\mu n_{1}\left|\xi\right|^2\nonumber\\
&&\left.-\frac{\mu m_{1}}{2}\xi^{\ast 2} -\frac{\mu m^{\ast}_{1}}{2}\xi^{2}\right\}\tilde{\sigma}(\xi^{\ast},\xi;t)\,.
\end{eqnarray}
The solution to Eq. (\ref{master2}) for the initial Gaussian state (\ref{gauss}) has the form
\begin{equation}\label{sol2}
\tilde{\sigma}(\xi^{\ast},\xi;t )=\exp\left[
-i\left(\xi^{\ast}\langle a\rangle_{t}+\xi \langle a\rangle_{t}\right)-
\frac{1}{2}\left(\xi^{\ast 2}\zeta(t)+\xi^{2}\zeta(t)^{\ast}\right)
-|\xi|^{2}\nu(t)\right]\,.
\end{equation}
The parameters of the state $\sigma_{t}$ appearing in (\ref{sol2}) satisfy the set of ordinary differential equations 
\begin{eqnarray*}
\dot{\langle a\rangle}_{t}&=&-\left(i\delta+\frac{\mu}{2}\right)\langle a\rangle_{t}-\sqrt{\mu}\beta_{1}\,,
\end{eqnarray*}
\begin{eqnarray*}
\dot{\zeta}(t)&=&-2i\delta\zeta(t)-\mu\left(\zeta(t)-m_{1}\right)\,,
\end{eqnarray*}
\begin{equation*}
\dot{\nu}(t)\;=\;-\mu\left(\nu(t)-n_{1}\right)\,.
\end{equation*}
The solution to this set of equations reads
\begin{eqnarray*}
\langle a\rangle_{t}&=&e^{-\left(i\delta+\frac{\mu}{2}\right)t}\left[\alpha_{0}-\frac{\sqrt{\mu} \beta_{1}}{i\delta+\frac{\mu}{2}}\left(e^{\left(i\delta+\frac{\mu}{2}\right)t}-1\right)
\right]\,,\\
\zeta(t)&=&e^{-\left(2i\delta +\mu\right)t}\left(\zeta_{0}-\frac{\mu m_{1}}{2i\delta +\mu}\right)+\frac{\mu m_{1}}{2i\delta+\mu}\,,\\
\nu(t)&=&e^{-\mu t}\left(\nu_{0}-n_{1}\right)+n_{1}\,.
\end{eqnarray*}

Let us notice that the asymptotic solutions have the form: $\zeta(\infty)=\frac{\mu m_{1}}{2i\delta+\mu}$, $\nu(\infty)=n_{1}$, and $\langle a\rangle_{\infty}\;=\;\frac{-\sqrt{\mu} \beta_{1}}{i\delta+\frac{\mu}{2}}$. 
It means that the systems approaches asymptotically the Gaussian state with parameters independent of their initial values. It should be noted that for the resonance case ($\delta=0$) the asymptotic {\it a priori} and {\it a posteriori} states coincide.

\end{document}